\pdfoutput=1
\documentclass[aps,pra,reprint,amsfonts,amsmath,showpacs,superscriptaddress,longbibliography]{revtex4-1}
\usepackage[english]{babel}
\usepackage[T1]{fontenc}
\usepackage{graphicx}
\usepackage{upgreek}

\renewcommand{\vec}[1]{\boldsymbol{#1}}
\newcommand\de{\mathrm{d}}
\newcommand\w{\omega}
\newcommand\wt{\tilde{\w}}
\newcommand\eps{\varepsilon}
\newcommand\rv{\vec{r}}
\newcommand\Ev{\vec{E}}
\newcommand\Evt{\tilde{\Ev}}
\newcommand\dg{^{\dagger}}
\newcommand\nr[1][]{_{\mathrm{nr}#1}}
\newcommand\II{\mathbb{I}}

\renewcommand\Im{\mathrm{Im}\,}
\renewcommand\Re{\mathrm{Re}\,}

\begin{document}
\title{Quasinormal-mode expansion of the scattering matrix}
\author{Filippo Alpeggiani}
\affiliation{Center for Nanophotonics, AMOLF, Science Park 104, 1098 XG Amsterdam, The Netherlands}
\affiliation{Kavli Institute of Nanoscience, Department of Quantum Nanoscience, Delft University of Technology, Lorentzweg 1, 2628 CJ Delft, The Netherlands}
\author{Nikhil Parappurath}
\author{Ewold Verhagen}
\affiliation{Center for Nanophotonics, AMOLF, Science Park 104, 1098 XG Amsterdam, The Netherlands}
\author{L. Kuipers}
\affiliation{Center for Nanophotonics, AMOLF, Science Park 104, 1098 XG Amsterdam, The Netherlands}
\affiliation{Kavli Institute of Nanoscience, Department of Quantum Nanoscience, Delft University of Technology, Lorentzweg 1, 2628 CJ Delft, The Netherlands}
\date{\today}

\begin{abstract}
It is well known that the quasinormal modes (or resonant states) of photonic structures can be associated with the poles of the scattering matrix of the system in the complex-frequency plane. In this work, the inverse problem, i.e., the reconstruction of the scattering matrix from the knowledge of the quasinormal modes, is addressed. We develop a general and scalable quasinormal-mode expansion of the scattering matrix, requiring only the complex eigenfrequencies and the \textit{far-field} properties of the eigenmodes. The theory is validated by applying it to illustrative nanophotonic systems with multiple overlapping electromagnetic modes. The examples demonstrate that our theory provides an accurate first-principle prediction of the scattering properties, without the need for postulating \textit{ad-hoc} nonresonant channels.
\end{abstract}

\maketitle

%%%%%%%%%%%%%%% Introduction %%%%%%%%%%%%%%%%%%%%%%%%%%%%%%%%%%%%%%%%%

Scattering matrices have been playing a ubiquitous role in physics since the early history of quantum field theory \cite{quantum}.
%In the very essence, the idea is to reduce the number of degrees of freedom of an open system by considering only a limited portion of the total phase space and assuming that the subsystem is coupled with the remaining environment by a finite number of ports or channels.
Nowadays, scattering-matrix techniques represent an irreplaceable tool for scientists working in nuclear physics \cite{nuclear}, electronic transport \cite{datta}, or classically chaotic systems \cite{chaos}, just to mention some of the several fields of application. Scattering matrices also enjoy a well deserved popularity in electromagnetic modeling, ranging from microwave devices \cite{microwave} to nanophotonics applications, such as scattering and transmission from nanostructured objects \cite{bohren,Smatrix2,liscidini2008}.

Most of the systems that are usually investigated with scattering-matrix techniques display a highly structured resonant response as a function of the excitation frequency (or energy), with the resonances in the spectrum being directly related to the poles of the analytical continuation of the scattering matrix in the complex-frequency plane \cite{popov1995,ishihara2002}. For electromagnetic systems, such poles correspond to quasinormal modes (also called resonant states), i.e., complex-frequency solutions of Maxwell's equations with outgoing-wave boundary conditions \cite{leung,quasinormal_muljarov1,quasinormal_hughes1,quasinormal_lalanne,quasinormal_muljarov2}. In a sense, quasinormal modes represent the bare skeleton around which the frequency-dependent response of the system is built. The interplay among different electromagnetic modes has proven to be crucial for explaining several intriguing phenomena, such as Fano resonances in optical systems \cite{fan2003}, scattering dark states \cite{soljacic2014,alaee2015}, and the optical analog of electromagnetically induced transparency and superscattering \cite{fan2012_prl}, and for designing new optical materials, such as optical metasurfaces for wavefront shaping \cite{staude_aom}. For these reasons, it is desirable and extremely interesting to be able to reconstruct \textit{ab initio} the entire scattering matrix of a system from the knowledge of its quasinormal modes. Not only would such quasinormal-mode expansion contribute to the understanding of complicated spectral features in terms of interference and superposition of resonant states, but it would also offer practical advantages from the numerical point of view, since a full eigenmode calculation is generally faster and more comprehensive than a large number of single-frequency simulations.

A promising theoretical platform in which to carry out this program is represented by temporal coupled-mode theory for optical resonators. Such framework has been fruitfully employed to study the transmission of layered photonic-crystal structures \cite{fan2002,fan2003,fan2013}, gratings \cite{bykov}, coupled cavities and waveguides \cite{fan2004,vuckovic}, and the scattering cross section of nanoparticles \cite{soljacic2007,fan2012,soljacic2014}. For the moment, however, coupled-mode theory has been typically restricted to a selection of only one or two modes of the optical system. The residual spectral response is accounted for by a slowly varying frequency-dependent background, which is typically fitted from simulation data \cite{fan2002,fan2003,fan2004,fan2013}. Part of the difficulty in expanding coupled-mode theory by including an arbitrary number of modes lies in estimating the coupling coefficients that relate the resonant states with the input--output channels. For a small number of modes, these can be obtained from symmetry considerations \cite{fan2003,fan2004}  or from the temporal decay rates \cite{fan2013}. However, in order to address the general case of multiple modes and an arbitrary configuration of input--output channels, a direct connection between the parameters of coupled-mode theory and the far-field properties of quasinormal modes is required.

In this work, we establish such a connection and we present a general theory to expand the scattering matrix on the quasinormal modes of photonic systems, which can be directly scaled to any number of eigenmodes and incoming or outgoing channels. The theory, based on the far-field asymptotic behavior of the modes and the unitarity property of the scattering matrix, represents a fully predictive tool that does not require the fitting of an additional nonresonant background.
%We validate our formalism by comparing its predictions with independent numerical simulations and obtaining excellent agreement between the two.
There are formal similarities between our results and the  expansion of the electromagnetic Green function on normalized quasinormal modes, which is a well known result from classical electrodynamics  \cite{leung,quasinormal_hughes2,lalanne2}; of course, when the expansion of the Green function is known for any point in space, then the scattering properties of the system can also be obtained \cite{lalanne2,commandre}. The theory that we present is formulated in a basis of input and output channels and it differs from these approaches in requiring only the far-field behavior of the modes at the input--output ports, as opposed to the full spatial distribution of the eigenfield. Moreover, our theory is independent of the choice of the normalization of the quasinormal modes.

Modal methods offer a deeper physical insight into the properties of resonant systems, because they allow us to draw a connection between the origin of complicated spectral features and the characteristics of the underlying quasinormal modes. For these reasons, they are particularly suitable for describing, understanding, and optimizing complex photonic systems. Notably, since the formalism that we present is derived on the basis of general coupled-mode theory, its range of applicability goes beyond that of classical electrodynamics.

The work is organized as follows. In Sec.~\ref{theory} we derive the quasinormal-mode expansion of the scattering matrix, whereas in Sec.~\ref{applications} we numerically validate the theory in the illustrative cases of photonic crystal slabs and multilayered metallic nanoparticles.

%%%%%%%%%%%%%%%%%% Theory %%%%%%%%%%%%%%%%%%%%%%%%%%%%%%%%%%%%%%%%%%%%%
\section{Theory}\label{theory}

\subsection{Quasinormal modes}\label{quasinormal}

In order to provide a rigorous motivation for the application of the formalism of coupled-mode theory to optical systems, we begin our analysis by establishing a direct connection with the theory of quasinormal modes.
We consider a system of dielectric or absorbing photonic structures, described by a spatially inhomogeneous distribution of the dielectric function $\eps(\rv, \w)$. We assume that in the limit $r \to \infty$, the dielectric function $\eps(\rv,\w)$ tends to the constant value $\eps_b$ and we define $\Delta\eps(\rv,\w) = \eps(\rv,\w) - \eps_b$ \footnote{The discussion in this section can be generalized to more complex situations (for instance, different values of the background dielectric constant in the two half spaces), provided that the corresponding dyadic Green tensor is used in Eq.~(4).}. The system supports a discrete number of quasinormal modes (also called resonant states) which are defined as the transverse complex-frequency solutions $(\Evt_j,\tilde{\vec{H}}_j)$ of Maxwell's equations,
\begin{equation}\label{maxwell_pair}
\begin{aligned}
-\frac{i}{\mu_0}\nabla \times \Evt_j(\rv) &= \wt_j \tilde{\vec{H}}_j(\rv), \\
\frac{i}{\eps_0\eps(\rv,\wt_j)}\nabla \times \tilde{\vec{H}}_j(\rv) &= \wt_j \Evt_j(\rv),
\end{aligned}
\end{equation}
with outgoing radiation boundary conditions \cite{quasinormal_muljarov1,quasinormal_hughes1,quasinormal_lalanne,quasinormal_muljarov2}. This linear system of equation is equivalent to a quadratic eigenproblem for the electric field:
\begin{equation}\label{maxwell}
\nabla \times \nabla \times \Evt_j(\rv) - \eps(\rv,\wt_j)\frac{\wt_j^2}{c^2}\Evt_j(\rv) = 0.
\end{equation}
As a consequence of the complex eigenfrequency $\wt_j$, quasinormal modes are characterized by a diverging amplitude in the far field.

The same considerations also apply to systems that are periodic in one or two dimensions and radiating in the remaining dimensions. In this case, the one- or two-dimensional crystalline momentum $\vec{k}$ is conserved and it is possible to define a family of quasinormal modes of the form:
\begin{equation}
\Evt_{\vec{k},j}(\rv) = e^{i \vec{k}\cdot\rv} \mathcal{E}_{\vec{k},j}(\rv),
\end{equation}
where the function $\mathcal{E}_{\vec{k},j}(\rv)$ has the same periodicity of the system and the field satisfies the outgoing radiation boundary conditions along the nonperiodic dimensions. To keep the notation general, we will assume the reciprocal wavevector to be fixed and omit the index $\vec{k}$.

For one-dimensional dielectric media and three-dimensional spheres, it has been proven that the modal eigenfields form a complete basis inside the structure, i.e., in the region where $\Delta\eps(\rv,\w) \ne 0$, provided that $\Delta\eps(\rv,\w)$ or any order of its derivative is discontinuous at the boundary of its domain \cite{leung}. In this work, we make the assumption that the completeness hypothesis holds for arbitrary resonant systems, as well \cite{quasinormal_hughes2,quasinormal_muljarov2}.

Following the usual scattering theory, we suppose that the system is illuminated by an incident field $\Ev_b$, which, in turn, is a solution of the wave equation \eqref{maxwell} with only the background dielectric constant, $\eps_b$. Splitting the total field in the incident and scattered components, $\Ev(\rv)= \Ev_b(\rv) + \Ev_s(\rv)$, the latter can be shown to satisfy the inhomogenous wave equation in the presence of a source term proportional to the incident radiation, i.e.,
\begin{equation}\label{scattering_eq}
\nabla \times \nabla \times \Ev_s(\rv) - \eps(\rv,\w)\frac{\w^2}{c^2}\Ev_s(\rv) = \Delta\eps(\rv,\w)\frac{\w^2}{c^2}\Ev_b(\rv).
\end{equation}
Limiting ourselves to the region where $\Delta\eps \neq 0$, in the assumption that quasinormal modes $\Evt_j$ form a complete basis, we can expand the scattered field on them:
\begin{equation}\label{E_exp_qnm}
\Ev_s(\rv) = \sum_j a_j \Evt_j(\rv).
\end{equation}
The exact expression for the coefficients $a_j$ depends on the incident field.
Eventually, the knowledge of $\Ev_s$ in a finite region is sufficient to extract the far-field properties of the scattered field, as it is described by the same Eq.~\eqref{scattering_eq}, which becomes
\begin{equation*}
\nabla \times \nabla \times \Ev_s(\rv) - \eps_b\frac{\w^2}{c^2}\Ev_s(\rv) = \Delta\eps(\rv,\w)\frac{\w^2}{c^2}\left[\Ev_b(\rv)+\Ev_s(\rv)\right].
\end{equation*}
This equation has the formal solution
\begin{multline}\label{scattering_eq2}
\Ev_s(\rv) = \int \de^3\rv' \Bigg [\Delta\eps(\rv',\w) \frac{\w^2}{c^2}
[\Ev_b(\rv') \\
+
\left. \sum_j a_j \Evt_j(\rv')]
\tilde{\vec{G}}_b(\rv,\rv',\w)\right],
\end{multline}
with $\tilde{\vec{G}}_b(\rv,\rv',\w)$ being the dyadic Green tensor of the background electromagnetic environment with homogeneous dielectric constant $\eps_b$. Since the integral in Eq.~\eqref{scattering_eq2} is limited to the region where $\Delta\eps \ne 0$, we were able to replace the field expansion of Eq.~\eqref{E_exp_qnm}.

At this point, we expand the input field over a set of incoming waves (or, more generally, ports), $\Ev_b(\rv) = \sum_{\alpha} s_{+\alpha} \Ev^{(+)}_{\alpha}$, and total electric field over a corresponding set of outgoing waves, $\Ev = \sum_{\alpha} s_{-\alpha} \Ev^{(-)}_{\alpha}$, whose detailed expression depends on the specific geometry of the system. Equation~\eqref{scattering_eq2} clearly shows that the amplitude of each outgoing wave, $s_{-\alpha}$, can be written as the sum of a direct channel, which is directly proportional to the incoming amplitudes $s_{+\alpha}$, and a resonance-mediated channel, which is proportional to the quasinormal-mode amplitudes $a_j$. In turn, the latter amplitudes are related to the incoming field through Eq.~\eqref{scattering_eq}. From the linearity of Maxwell equations, it follows that all these relations can be written in terms of linear operators. This is the basis of the coupled-mode formalism, which we illustrate in the following.

\subsection{Coupled-mode equations}

\begin{figure}[t]
\includegraphics{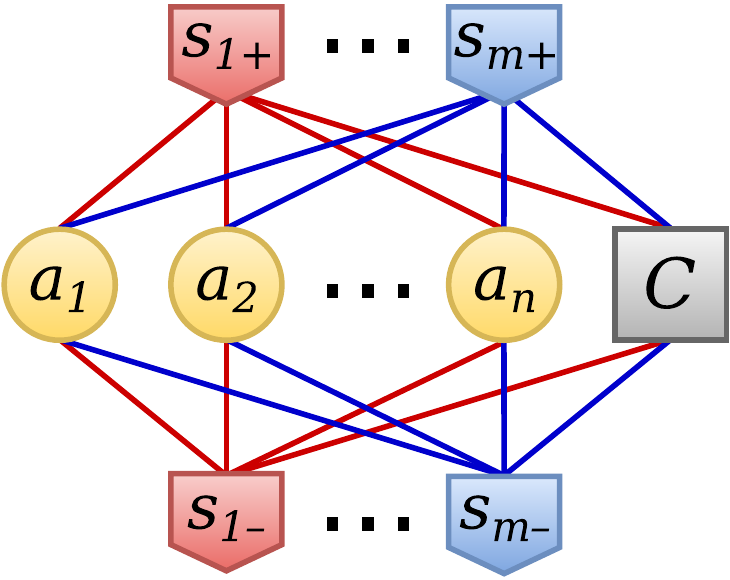}
\caption{(Color online) Schematic of $m$ ports coupled to $n$ quasinormal modes with amplitudes $a_j$ ($j = 1, \dots, n$) and linked by a direct-coupling term $C$. The notation $s_{p+}$ and $s_{p-}$ ($p = 1, \dots, m$) refers to the amplitude of incoming and outgoing waves, respectively.}\label{scheme} 
\end{figure}

Seeking a more general formulation, we write the characteristic equation of quasinormal modes, Eq.~\eqref{maxwell_pair}, as an eigenvalue problem for the effective ``Hamiltonian'' $\Omega + i\Gamma$,
\begin{equation}\label{eigen}
(\Omega + i\Gamma) \vec{a}_j = \tilde{\omega}_j \vec{a}_j.
\end{equation}
Here and in the following, we assume the convention $\exp(i\w t)$ for the temporal dependence of the field. The components of the vectors $\vec{a}_j$ are interpreted as the coefficients of the expansion of the electric field in terms of quasinormal modes, according to Eq.~\eqref{E_exp_qnm}. Due the inherently dissipative nature of quasinormal modes, the Hamiltonian operator $\Omega + i\Gamma$ is non-Hermitian and it has been split in the Hermitian and skew-Hermitian parts, which are expressed in terms of the two Hermitian operators $\Omega$ and $\Gamma$. Using the same language of Eq.~\eqref{eigen} and following our previous considerations, we relate the incoming and outgoing amplitudes of the electromagnetic field (which we express in vector form as $\vec{s}_{+}$ and $\vec{s}_{-}$) by means of a system of coupled-mode equations:
\begin{align}\label{fan1}
i\omega \vec{a} &= i(\Omega + i\Gamma) \vec{a} + K^T \vec{s}_+
\\\label{fan2}
\vec{s}_- &= C\vec{s}_+ + D\vec{a}.
\end{align}
The operator $C$ represents the direct-coupling channel, whereas the operators $K$ and $D$ account for coupling between quasinormal modes and the incoming and outgoing ports, respectively. Although there might be in principle infinitely many quasinormal modes and ports, for practical reasons we assume that the number of modes and ports is truncated to the finite values $n$ and $m$, respectively. In this way, all the operators reduce to finite-size matrices. The set of Eqs.~\eqref{fan1} and \eqref{fan2} is summarized by the scheme in Fig.~\ref{scheme}.

As originally demonstrated in Refs.~\cite{fan2003} and \cite{fan2004}, some relations among the quantities that appear in Eqs.~\eqref{fan1} and \eqref{fan2} can be directly deduced from some very general physical properties of the system. First, electromagnetic reciprocity and energy conservation imply that
\begin{equation}\label{fan2half}
K = D
\end{equation}
and
\begin{equation}\label{fan3}
\Gamma = \tfrac{1}{2}D\dg D + \Gamma\nr,
\end{equation}
respectively. In Eq.~\eqref{fan3}, we have straightforwardly extended the theory to include the (Hermitian) decay matrix $\Gamma\nr$, which accounts for absorption and other potential nonradiative-dissipation channels.
Moreover, by comparing the dynamics described by Eqs. \eqref{fan1} and \eqref{fan2} with the time-reversed case and employing time-reversal symmetry, it can be shown that
\begin{equation}\label{fan4}
CD^* = -D.
\end{equation}

The system in Eqs.~\eqref{fan1} and \eqref{fan2} has been extensively used to model the scattering properties of various photonic structures \cite{fan2002,fan2004,fan2012,fan2013,soljacic2014}, proving itself particularly valuable for investigating the physical mechanism at the basis of various phenomena, such as the formation of Fano lineshapes in the spectrum as a consequence of the interference between the resonant and the  direct-coupling channels \cite{fan2003}. In all these cases, however, the number of modes included in the equations is limited to one or two, and the direct-coupling channel, if present, is accounted 
for by fitting a specific frequency-dependent background response obtained from independent numerical simulations of the spectrum (see, for instance, Refs.~\cite{fan2003,fan2004,fan2012}). The need for independent frequency-by-frequency simulations restricts the suitability of coupled-mode theory as a first-principle computational tool. Moreover, accurately fitting the direct-coupling background typically requires some additional assumptions which are difficult to interpret on physical grounds (for instance, the need for a frequency-dependent effective dielectric constant). In a broader sense, the actual separation between the resonant states and the frequency-dependent background is somewhat arbitrary, since the latter is also made up of a number of broad resonances associated with additional quasinormal modes. In the light of our assumption about the completeness of quasinormal modes (Sec.~\ref{quasinormal}), we expect that by enlarging the set of electromagnetic modes, so to include the resonances usually associated with the background, we could remove the need for fitting the direct-coupling background and treat all resonant states on equal grounds. In this way, in addition to getting a more transparent physical picture, we could also better elucidate the modal structure at the basis of resonant systems. Implementing this strategy represents one of the main motivations for the formalism that we present in the next section, which is easily scalable to multiple modes with varying decay rates.

\subsection{Expansion of the scattering matrix}\label{sec:expansion}

The scattering matrix of the system connects the amplitude of the outgoing waves with the amplitude of the incoming waves:
\begin{equation}\label{S}
S = C - i D(\w\II-\Omega-i\Gamma)^{-1}D^T,
\end{equation}
where we used identity \eqref{fan2half}. Here, we derive an expression for the expansion of the scattering matrix on quasinormal modes, on the basis of the system of Eqs.~\eqref{fan1}--\eqref{fan2}.

To this purpose, in addition to the complex eigenfrequencies of the quasinormal modes, $\wt_j$ ($j = 1,\dots,n$), we also assume the knowledge of the asymptotic behavior of the quasinormal-mode eigenfield in the output ports, which is equivalent to the knowledge of the relative complex amplitudes of the vectors
\begin{equation}\label{eigen2}
\vec{b}_j \doteq \left.\vec{s}_-\right|_{\w = \wt_j} = D\vec{a}_j.
\end{equation}
For simplicity, we will refer to the vectors $\vec{b}_j$ as the ``scattering eigenvectors'' of the system.  As it is the case for all eigenproblems, the (complex) normalization constant of the eigenvectors can be set arbitrarily; however, as proven in App.~\ref{mult}, the final expression for the scattering matrix does not depend on the choice of such constant. As a consequence, our approach is inherently normalization-free, at variance with other works dealing with the expansion of the dyadic Green function, which require the quasinormal modes to be normalized in a specific fashion \cite{quasinormal_hughes1,quasinormal_lalanne,quasinormal_muljarov2}.

In practice, the complex eigenvalues and the scattering eigenvectors need to be computed by numerical eigensolvers. The specific method depends on the definition of the input--output ports, but, in general, it involves calculating the electric field at a point or on a surface in the far-field region of the system, and, possibly, computing the projection integral of the field with the modal profile of the port. Some examples are provided in Sec. II. We stress that, since the scattering eigenvectors depend only on the far-field behavior of the resonant states, they can be obtained without computing the full distribution of the electromagnetic field over all space. This characteristic is particularly helpful, for instance, when quasinormal modes are calculated with numerical techniques such as the boundary-element method or the multipole expansion method, which typically benefit from a faster rate of convergence for far-field calculations.

Since the matrix $\Omega + i\Gamma$ is not Hermitian, the right eigenvectors alone are not orthogonal. However, as it is known from the theory of complex Hamiltonians \cite{complex}, right eigenvectors ($\vec{a}_j$ in our case) form a biorthogonal basis together with left eigenvectors, which are defined by the equation
\begin{equation}
\vec{l}_j\dg (\Omega + i\Gamma) = \wt_j \vec{l}_j\dg.
\end{equation}
To simplify the notation, we introduce  the $n \times n$ matrix $A$ whose columns are the right eigenvectors $\vec{a}_j$ and the corresponding matrix $L$ of the column left eigenvectors $\vec{l}_j$. With this new notation, Eq. \eqref{eigen} becomes:
\begin{equation}\label{eigen3}
(\Omega + i\Gamma) A = A \tilde{\Omega},
\end{equation}
with $\tilde{\Omega}$ being the diagonal matrix of the complex eigenvalues $\wt_j$. Moreover, we  define the $m \times n$ matrix $B$ whose columns are the vectors $\vec{b}_j$.

The complex Hamiltonian of Eq.~\eqref{eigen} can then be expanded on the biorthogonal basis as follows \cite{complex}:
\begin{equation}\label{expansion}
\omega\mathbb{I} - \Omega - i\Gamma = A(\omega\mathbb{I} - \tilde{\Omega})L\dg.
\end{equation}
Even if the right eigenvectors are not orthogonal, they are however linearly independent \cite{complex}; thus, we can formally write $L = (A\dg)^{-1}$. Replacing Eq.~\eqref{expansion} into Eq.~\eqref{S}, we obtain the quasinormal-mode expansion of the scattering matrix,
\begin{equation}\label{main}
S = C - i B \frac{1}{\omega\mathbb{I} -\tilde{\Omega}} \Lambda^{-1} B^T,
\end{equation}
where we define $\Lambda \doteq A^T A$ and we use the relation $B = DA$, which comes directly from Eq.~\eqref{eigen2}. For the moment, Eq.~\eqref{main} represents only a formal result, which can be also  seen as a special case of Mittag-Leffler's theorem on the pole-expansion of meromorphic functions \cite{mittag}. For all practical purposes, it is crucial to derive an expression for the matrix $\Lambda$. This latter matrix plays a fundamental physical role, because the amplitude and phase of its terms determine the oscillator strength of each resonance and affect the degree of interference among the modes, which, in turn, has been found responsible for the appearance of interesting spectral features, such as Fano lineshapes \cite{fan2003} or the optical analogue of electromagnetically-induced transparency \cite{fan2012_prl}.

First of all, it can be shown that $\Lambda$ is diagonal. This result follows from the symmetry of the complex Hamiltonian $\Omega + i\Gamma$, which can be proven by combining Eqs.~\eqref{fan3} and \eqref{fan4}. The same result can also be derived from the requirement that the resulting scattering matrix must be symmetric \cite{fan2004}. Next, by multiplying each side of Eq.~\eqref{fan4} by $A^*$ and after some algebraic manipulations, we obtain
$CB^* = -B \Lambda^{-1} (A\dg A)^*$, which we can recast in the more compact form
\begin{equation}\label{main2}
C B^* + B \Lambda^{-1} Q^* = 0,
\end{equation}
which defines the matrix
$
Q = A\dg A.
$

By multiplying Eq.~\eqref{eigen3} by $A\dg$ on the left, taking the difference with its Hermitian conjugate, and employing Eqs.~\eqref{fan3} and \eqref{eigen2}, we arrive at
\begin{equation}\label{Qraw}
Q \tilde{\Omega}
 - \tilde{\Omega}^* Q = 2i A\dg \Gamma A = i B\dg B + 2iA\dg \Gamma\nr A.
\end{equation}
In general, the solution for $Q$ cannot be written explicitly in terms of matrix products; however, it is straightforward to express it componentwise. First, in the case of no absorption ($\Gamma\nr = 0$), we can write:
\begin{equation}\label{Q}
Q_{ij}
= i\frac{\vec{b}_i\dg \vec{b}_j}{\wt_j - \wt_i^*}.
\end{equation}
This latter equation allows us to clarify the physical meaning of Eq.~\eqref{main2}. With the aid of Eqs.~\eqref{main} and \eqref{Q}, it can be shown that Eq.~\eqref{main2} is equivalent to the condition
%In fact, after replacing Eq.~\eqref{Q}, Eq.~\eqref{main2} reads columnwise
%\begin{equation}
%\left[C - i \vec{b}_i \frac{1}{\wt_j^* - \wt_i} \frac{1}{\lambda_i^*} \vec{b}_i^T\right]
%\vec{b}_j^* = \vec{0},
%\end{equation}
%which, taking into account Eq.~\eqref{main} and the symmetry of the scattering matrix, is %equivalent to assuming
\begin{equation}\label{unitarity}
S\dg(\wt_j) \vec{b}_j = \vec{0}.
\end{equation}
From the inversion of the scattering matrix, on the other hand, we obtain that $S^{-1}(\wt_j)\vec{b}_j = 0$, since quasinormal modes are defined as the self-sustaining solution of Maxwell's equations in the absence of any input radiation. Comparing the two results, it is clear that Eq.~\eqref{main2} guarantees that the scattering matrix is unitary at the modal eigenfrequencies, as required by energy conservation.

In the presence of absorption ($\Gamma\nr \ne 0$), the energy balance must account also for the additional dissipation. In a broad sense, nonradiative processes represent a number of input--output channels that it is impractical to take into account directly. It is possible, however, to quantify their total effect on the decay rate of each quasinormal mode, for instance by calculating the shift of the imaginary part of the eigenfrequency with respect to the case when all losses are turned off. An example of this approach is discussed in Sec.~IID. When the nonradiative decay rate is small compared to the frequency of the mode, the nonradiative term $\Gamma\nr$ can be treated as a first-order perturbation of the total Hamiltonian, i.e., we can assume $A\dg \Gamma\nr A \simeq A\dg A \tilde{\Gamma}\nr = \tilde{\Gamma}\nr A\dg A$, where $\tilde{\Gamma}\nr$ is the diagonal matrix of the first-order nonradiative decay rates, $\gamma\nr[,j]$ ($j = 1,\dots,n$). In this way, we can write the following generalized expression for $Q$:
\begin{equation}\label{Qnr}
Q_{ij}
= i\frac{\vec{b}_i\dg \vec{b}_j}{\wt_j - i\gamma\nr[,j]
 - \wt_i^* - i\gamma\nr[,i]}.
\end{equation}

Equations~ \eqref{main2} and \eqref{Q} [or \eqref{Qnr} for absorbing systems] allow us to fully determine the matrix $\Lambda$, and, hence, the quasinormal-mode expansion of Eq.~\eqref{main}. However, a closer inspection of Eq.~\eqref{main2} reveals that the system has $m \times n$ equations (the dimension of $B$) and only $n$ unknows (the diagonal of $\Lambda$). Thus, for a given direct coupling matrix $C$, the system is generally overdetermined and a solution is not always guaranteed to exist. From a different perspective, the direct matrix $C$ cannot be chosen freely, but it must satisfy some constraints that depend on the properties of the resonant states. In practice, it might be difficult to choose a direct matrix with a simple analytical form and, at the same time, consistent with Eq.~\eqref{main2}, especially when a large number of quasinormal modes is involved.

For all these reasons, it is essential to develop a general theory that encompasses also the case when the matrix $C$ is an approximation of the exact direct-coupling matrix. To this end, instead of looking for an exact solution of Eq.~\eqref{main2}, we search for an approximate solution in the least-square sense. To be more precise, having defined the vectors $\vec{x}_j$ ($j = 1,\dots,n$) as the columns of the matrix
\begin{equation}\label{X}
X = C B^* (Q^*)^{-1},
\end{equation}
we look for the diagonal matrix $\Lambda$ in Eq.~\eqref{main2} whose diagonal terms, $\lambda_j$, minimize the objective function
\begin{equation}
f(\lambda_1,\dots,\lambda_n) = \sum_{j=1}^{n} | \lambda_j \vec{x}_j + \vec{b}_j |^2.
\end{equation}
This reformulation of the problem does not affect the generality of the theory, because, if Eq.~\eqref{main2} has an exact solution, then such solution must coincide with the least-square one \cite{algebra}.

A simple calculation of the stationary points of the objective function leads to the result $\lambda_j = -\vec{x}_j\dg \vec{b}_j / (\vec{x}_j\dg \vec{x}_j)$, which, once replaced into Eq.~\eqref{main}, provides us with the final expression
\begin{equation}\label{main3}
S = C + i\sum_{j=1}^{n}
\frac{\vec{x}_j\dg \vec{x}_j}{\vec{x}_j\dg \vec{b}_j}
\frac{\vec{b}_j \vec{b}_j^T}{\omega - \wt_j}.
\end{equation}
Equation~\eqref{main3}, together with Eqs.~\eqref{X} and \eqref{Q}, is the desired expansion of the scattering matrix and it represents the main result of the present work. Using Eq.~\eqref{X}, the expansion coefficients can be explicitly written as
\begin{equation}
\frac{1}{\lambda_j} = -\frac{\vec{x}_j\dg \vec{x}_j}{\vec{x}_j\dg \vec{b}_j}
= -\frac{\sum_{nn'}Q^{-1}_{nj}\left(Q^{-1}_{n'j}\right)^* \vec{b}^T_{n} C\dg C \vec{b}^*_{n'}}{\sum_n Q^{-1}_{nj} \vec{b}^T_n C\dg \vec{b}_j}.
\end{equation}
The denominator of the coefficient can be regarded as a modified inner product that renormalizes the  scattering eigenvectors in order to guarantee the total scattering matrix to be unitary. In the limiting case when the offdiagonal elements of $Q$ are negligible, the expression in Eq.~\eqref{main3} reduces to a modified version of the prominent Breit-Wigner formula of nuclear physics  \cite{nuclear,chaos}, as shown in App.~\ref{nondiagonal}. In addition, in App.~\ref{mult} we also show that the result in Eq.~\eqref{main3} is independent of the normalization of the scattering eigenmodes.

\section{Applications}\label{applications}

\subsection{Photonic crystal slab}\label{sec:PhC}

\begin{figure}
\includegraphics{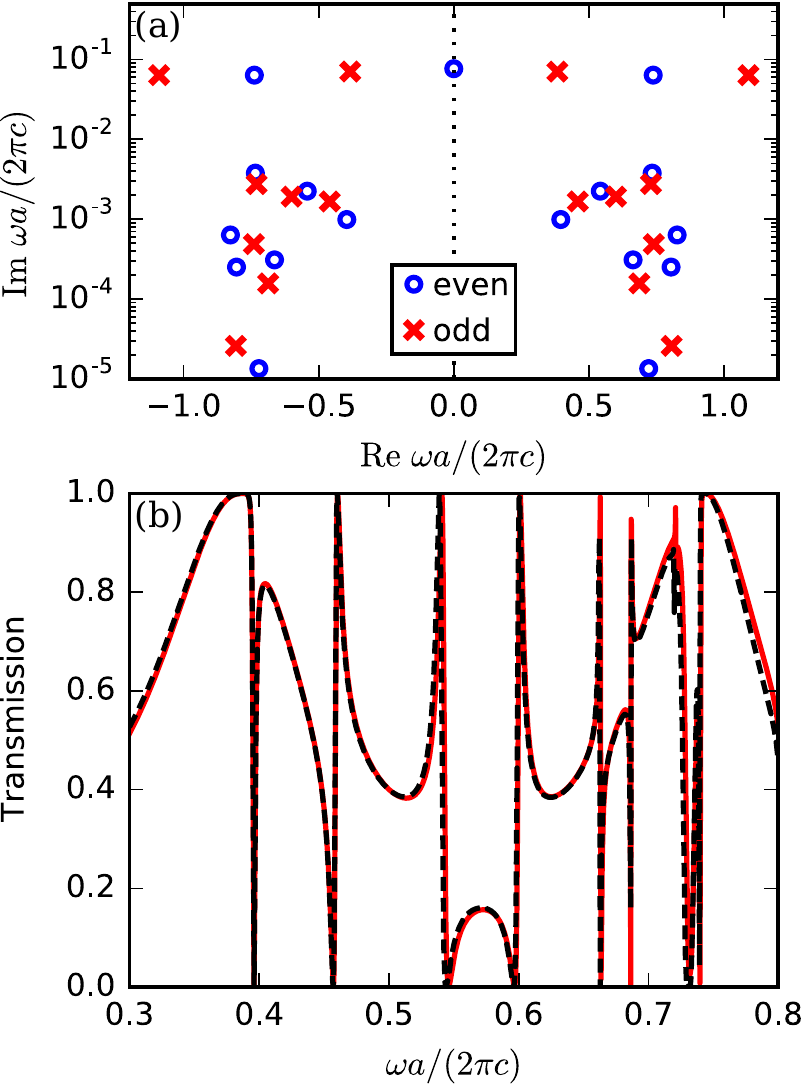}
\caption{(Color online) Application of the theory to a photonic crystal slab composed of a square lattice of air holes etched in a suspended silicon membrane. (a) Real and imaginary part (log scale) of the quasinormal-mode complex eigenfrequencies, together with the corresponding symmetry of the modes (even, odd) by inversion with respect to the slab middle plane. (b) Transmission intensity computed by expanding the scattering matrix on the quasinormal modes (red solid line), compared with the exact result by the Fourier modal method (dashed line) \cite{s4}.}\label{PhC} 
\end{figure}

As an illustrative example, we consider a photonic crystal slab composed of a square lattice of circular holes etched in a silicon membrane ($\eps = 12.1$). Indicating with $a$ the lattice constant, we assume the slab thickness and the hole radius $t = 0.4a$ and $r = 0.2a$, respectively. For normally incident light polarized along one of the lattice axes, we can limit ourselves to a single polarization of light; moreover, in the range of frequency $\w < 2\pi/a$, only the zeroth order of diffraction is  available. As a consequence, the system can be effectively described with two ports, corresponding to the plane waves $E_{1,+} = s_{1,+} e^{-i \w z / c}$ and $E_{2,+} = s_{2,+} e^{i \w z / c}$, propagating along the normal direction to the slab, which we indicate as the $z$ axis.

In Fig.~\ref{PhC}(a) we show the complex eigenfrequencies of the quasinormal modes of the system for normally incident light. Although all the modes represent equally valid solutions of the same characteristic equation \eqref{maxwell} and they are treated on equal grounds in the expansion of the scattering matrix \eqref{main3}, it is useful from a physical point of view to distinguish between two categories of quasinormal modes: weakly dissipating quasi-guided modes and leaky modes with much larger radiation rates. As it appears from Fig.~\ref{PhC}(a), the threshold between the two families can be set around $\Im\wt \approx 10^{-2}\,(2 \pi c /a)$, with a difference of more than one order of magnitude between the corresponding imaginary parts of the eigenfrequencies. The leaky modes ($\Im\wt > 10^{-2}\, 2 \pi c /a$) have strong similarities with the Fabry-P\'{e}rot resonances of a homogeneous dielectric slab with an average refractive index $n_{\mathrm{av}}$, displaying a roughly constant frequency spacing of the order of the free spectral range $\delta\w = \pi c/(n_{\mathrm{av}} t)$. The deviation from the equal spacing behavior grows when the frequency increases, due to the wavelength becoming more sensitive to the dielectric-function inhomogeneity in the system \cite{fan2002}.

Quasi-guided modes can be easily computed in various ways, including, e.g., frequency-domain \cite{quasinormal_lalanne} or time-domain \cite{quasinormal_hughes1,quasinormal_hughes2} methods, or by determining the poles of the scattering or transmission coefficient in the complex frequency plane \cite{popov1995}. These techniques can also be combined, in order to exploit specific advantages. For instance, in the present example, the modes with $\Re\wt_j > 0$ have been computed by solving a linearized version of the eigenproblem in Eq.~\eqref{maxwell} with a commercial finite-element package \cite{comsol}, whereas, for better numerical accuracy, leaky modes have been obtained separately by looking for the complex-frequency poles of the transmission amplitude computed with the Fourier modal method using a freely available solver \cite{s4}. Since the wave equation \eqref{maxwell} is second order in the frequency, for each quasinormal mode with $\Re\wt_j > 0$ there exists a corresponding state with $\wt_{j'} = -\wt_j^*$ and $\Evt_{j'}(\vec{r}) = \Evt_j^*(\vec{r})$ \cite{leung}, which has been included in the calculations,  raising the total number of quasinormal modes under consideration in this example to $n = 33$. Due to numerical difficulties in performing the calculations near the imaginary axis of the complex-frequency plane, the decay rate of the $\Re\wt_j = 0$ mode has been estimated using the analytical formula for a homogeneous dielectric slab with an averaged refractive index \cite{quasinormal_muljarov1}.

The asymptotic behavior of the eigenfield is entirely determined by the inversion symmetry of the system with respect to the middle plane of the slab. Since the electric field amplitude is either even or odd with respect to the inversion, as indicated in Fig.~\ref{PhC}(a), we can directly assume the scattering eigenvectors
\begin{equation}\label{PhC_b}
\vec{b}_{\pm} = \frac{1}{\sqrt{2}} \left[\begin{array}{c}1 \\ \pm 1 \end{array}\right].
\end{equation}
for even (``$+$'') and odd (``$-$'') modes. As we already remarked, since the scattering matrix expansion is independent of the eigenfield normalization, any other choice of the normalization in Eq.~\eqref{PhC_b} would have been equally suitable. Finally, in agreement with our assumption about the completeness of quasinormal modes for photonic systems, we take the $2\times 2$ identity matrix as the direct-coupling matrix
\begin{equation}
C = \mathbb{I}_{2\times 2}.
\end{equation}

In this way, we can derive the expression of the scattering matrix of the photonic crystal slab by applying Eq.~\eqref{main3} with the complex eigenfrequencies of Fig.~\ref{PhC}(a) and the scattering eigenvectors of Eq.~\eqref{PhC_b}. The transmission intensity obtained from the resulting scattering matrix is shown by the solid curve in Fig.~\ref{PhC}(b), and it is compared with an independent calculation by the Fourier modal method (dashed line) \cite{s4}. The agreement between the curves is excellent, highlighting the validity of the quasinormal-mode expansion of the scattering matrix. The comparison confirms that the first-principle description of the optical properties of the system provided by the theory is complete and accurate; moreover, we stress that such a description does not require any \textit{ad-hoc} assumptions on the direct coupling channel and is based only on the complex eigenfrequencies of the quasinormal modes.

\subsection{Asymmetric photonic crystal structure}

\begin{figure}[h!]
\includegraphics{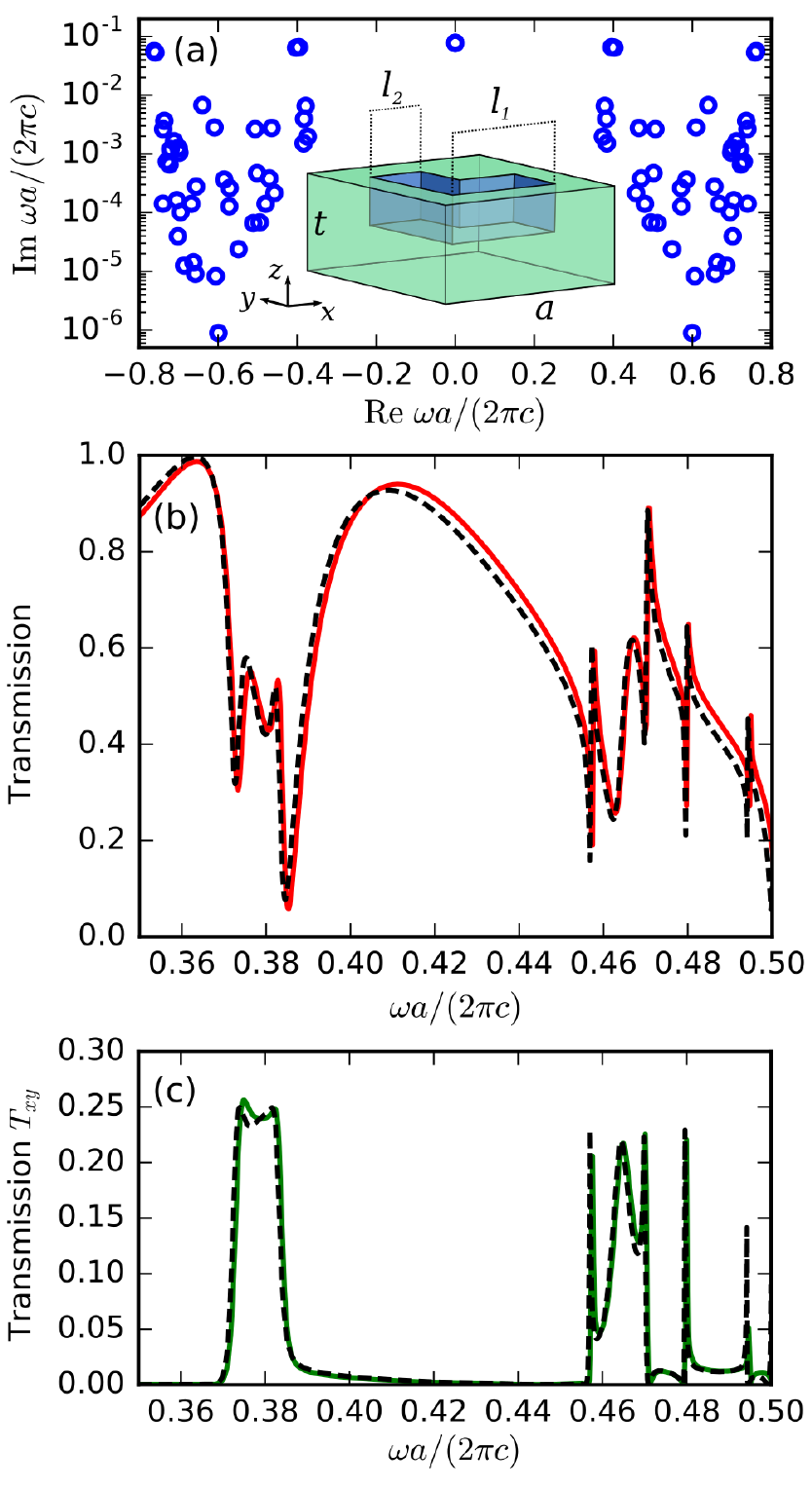}
\caption{(Color online) (a) Circles: real and imaginary part (log scale) of the quasinormal-mode eigenfrequencies of a square lattice of L-shaped patterned structures in a silicon membrane. The unit cell of the structure is represented in the inset ($t = 0.4a$, $l_1 = 0.6a$, and $l_2 = 0.3a$). Note that the structure is not symmetric by inversion along the $z$ axis. (b) Total transmittance, $T$, and (c) cross-polarized transmittance, $T_{xy}$, computed by expanding the scattering matrix on the quasinormal modes (solid line), compared with the exact result by the Fourier modal method (dashed line) \cite{s4}.}\label{fig:Lmodes} 
\end{figure}

A specific advantage of the scattering-matrix expansion is the straightforward applicability to generic systems lacking any particular symmetry. In order to illustrate this point, we consider a square lattice of L-shaped void structures partially patterned in a silicon slab. The shape and size of the structures is schematized in the inset of Fig.~\ref{fig:Lmodes}(a). The height of the patterned region ($h = 0.2a$) is one half of the total thickness of the slab ($t = 0.4a$), resulting in a configuration which is not symmetric by inversion along $z$. Moreover, for incident light polarized along one of the lattice main axes, the transmitted and reflected radiation will include a cross-polarized fraction. Therefore, we can model the system by defining four ports, corresponding to plane waves propagating above and below the slab and polarized along the two in-plane crystal axis (which we indicate as $x$ and $y$). In agreement with the assumption that quasinormal modes form a complete basis, we also assume the identity matrix as the direct-coupling matrix, i.e., $C = \mathbb{I}_{4\times 4}$.

The complex eigenfrequencies of the quasinormal modes, computed with the finite-element method \cite{comsol}, are presented in Fig.~\ref{fig:Lmodes}(a). Even in this case we can distinguish between a set of quasi-guided modes and a set of roughly equispaced leaky modes with a larger dissipation rate. Similarly to the previous example, the decay rate of the pair of modes with $\Re\wt_j = 0$ has been estimated using the analytical results for a homogeneous dielectric slab, and, moreover, we have also explicitly included the modes with $\tilde{\w}_{j'} = -\tilde{\w}_j^*$. However, in this case the scattering eigenvectors $\vec{b}_j$ must be obtained from the asymptotic behavior of the calculated quasinormal-mode eigenfield \cite{supplementary}. To this purpose, we consider the $x$ and $y$ electric-field components of each quasinormal mode in two planes located above and below the silicon slab at a sufficiently large distance to make the near-field contributions negligible. The specific choice of the distance does not affect the results, since only the relative amplitudes among the field components are relevant for the theory. It is interesting to note that the scattering eigenvector can also be computed with a near-to-far-field transformation of the quasinormal modes \cite{lalanne_far_field}.

From the expansion of the scattering matrix in Eq.~\eqref{main3}, we computed the total transmission intensity, $T$, and the cross-polarized transmission intensity, $T_{xy}$ (i.e., intensity of $x$-polarized transmitted light for $y$-polarized incident radiation). These quantities are shown (solid curves) in Figs.~\ref{fig:Lmodes}(b)--(c) and they are compared with the exact results (dashed curves) obtained from the Fourier modal method \cite{s4}. There is good agreement between the curves, especially in the vicinity of multiple narrow resonances, further confirming the validity of our approach as a predictive tool for computing the scattering matrix of electromagnetic systems. The small deviation from the exact result in the high-frequency region of Fig.~\ref{fig:Lmodes}(b) is  likely due to the lower number of leaky modes included in this example with respect to the case of Sec.~\ref{sec:PhC}. The large radiative width of leaky modes (with a quality factor of the order of 10) implies that additional states beyond the frequency range under consideration may still have a small effect on the transmission in Fig.~\ref{fig:Lmodes}(b). To corroborate this hypothesis, we verified that the agreement with simulation data can be further improved when an additional pair of leaky modes at $\Re\wt_j \simeq 1.1 (2\pi c/a)$ is included in the scattering matrix expansion \cite{supplementary}.

Computing the eigenvalues of Fig.~\ref{fig:Lmodes}(a) with the finite-element method takes about one hour on a multiprocessor workstation. By comparison, on the same workstation the time required by a single frequency-point calculation of the transmission using the same finite-element solver and the same mesh is about three minutes, implying that computing the transmission spectrum of Fig.~\ref{fig:Lmodes}(b) (roughly 1000 points) would require about 50 hours with the finite-element method. This 50-fold reduction of computational time highlights the computational advantage of modal methods over frequency-domain full-wave simulations using the same electromagnetic solver.

\subsection{Hybrid plasmonic system}

\begin{figure}
\includegraphics{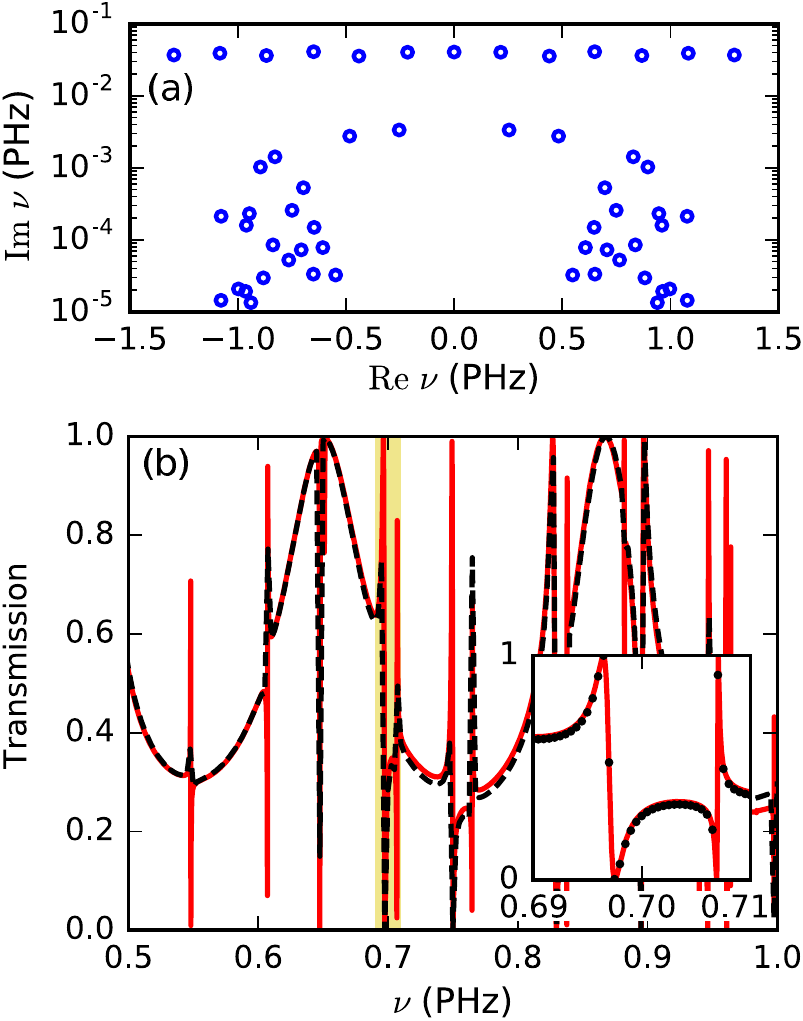}
\caption{(Color online) (a) Circles: real and imaginary part (log scale) of the complex eigenfrequencies of a square lattice (lattice constant $a = 200$ nm) of 60-nm-diameter metallic particles embedded in a 200-nm dielectric slab ($\eps = 12.1$). (b) Transmission intensity computed from the quasinormal-mode expansion of the scattering matrix (red solid line), compared with a finite-element frequency-by-frequency calculation (dashed line) \cite{comsol}. The inset shows a close-up of the spectrum in the highlighted frequency region. The dots indicate the frequency-by-frequency calculation.}\label{fig:hybrid} 
\end{figure}

Hybrid nanophotonic devices combining different photonic elements hold great promise for enhancing the functionality and the performance of various optical elements \cite{hybrid_review}. For instance, hybrid photonic--plasmonic systems made of plasmonic nanostructures coupled to optical resonators have been demonstrated to combine strong localization of light with precise control of the emission properties, enhancing the interaction with quantum emitters and the optical biosensing capabilities \cite{hybrid_chanda,hybrid_koenderink}.

In order to exemplify the applicability of our theory to this class of systems, we consider the example of a square array of 60-nm-diameter metallic particles embedded in  200-nm-thick dielectric slab. For simplicity, we assume the metal dielectric function to follow the dissipationless Drude's model, $\varepsilon(\w) = 1 - \w_p^2/\w^2$, with the plasma frequency $\hbar\w_p = 6 \mathrm{eV}$. The complex eigenfrequencies of the quasinormal modes have been computed with the finite-element method and are shown in Fig.~\ref{fig:hybrid}(a). In addition to the Fabry-P\'{e}rot resonances of the slab, the calculation reveals the presence of a number of narrower modes. Such modes originate from the hybridization of the multipolar modes of the metallic particle due to the interaction with the polarizable dielectric. Similarly to the case of Sec.~\ref{applications}A, the modes can be classified as even or odd with respect to inversion symmetry along the direction perpendicular to the slab.

The transmission spectrum of the system is derived from the quasinormal-mode expansion of the scattering matrix [Eq.~\eqref{main3}] and it is shown by the solid curve in Fig.~\ref{fig:hybrid}(b). Like the previous examples, we include the Fabry-P\'{e}rot zero-frequency mode with $\mathrm{Re}\wt_j = 0$ and we assume the unitary direct coupling matrix $C = \mathbb{I}_{2\times 2}$. The presence of a large number of modes results in a highly-structured spectrum with several closely-spaced minima and maxima of transmission, resulting from the reciprocal interference of light scattered by the polarization currents in the metal and the dielectric.

The transmission computed from the modal expansion of the scattering matrix is in very good agreement with the results of a frequency-by-frequency calculation with the finite-element method [dashed curve in Fig.~\ref{fig:hybrid}]. The calculation of the complex eigenvalues in Fig.~\ref{fig:hybrid}(a) takes a few hours on a multiprocessor workstation, comparing very favourably with the frequency-by-frequency computation, which requires about 30 hours using the same mesh. Furthermore, the modal expansion of the scattering matrix allows us to accurately resolve even the narrowest resonances, as demonstrated by the inset of Fig.~\ref{fig:hybrid}(b), displaying a close-up of the spectrum in a small frequency range. This characteristic emphasizes an advantage of modal methods over the direct frequency-by-frequency computation, where a reduction of the frequency resolution over the whole extent would be highly impractical on grounds of the increased computational cost.

All these considerations can be directly extended to more realistic devices, such as metallic nanoparticles in interaction with large optical resonators and photonic cavities. In these cases, the systems are expected to benefit even further from the advantages of the modal expansion method, due to the increased size and complexity. Notably, the interest of determining the quasinormal modes is not limited to accessing the scattering properties of the system. For instance, it has been demonstrated that the quasinormal modes of an array of metallic particles can interact with molecular excitons, giving rise to plasmon--exciton--polaritons \cite{gomezrivas}. Thus, in addition to providing access to the scattering matrix, the modal information is also essential for describing and understanding the polaritonic effects.

\subsection{Layered metallic particle}

\begin{figure}
\includegraphics{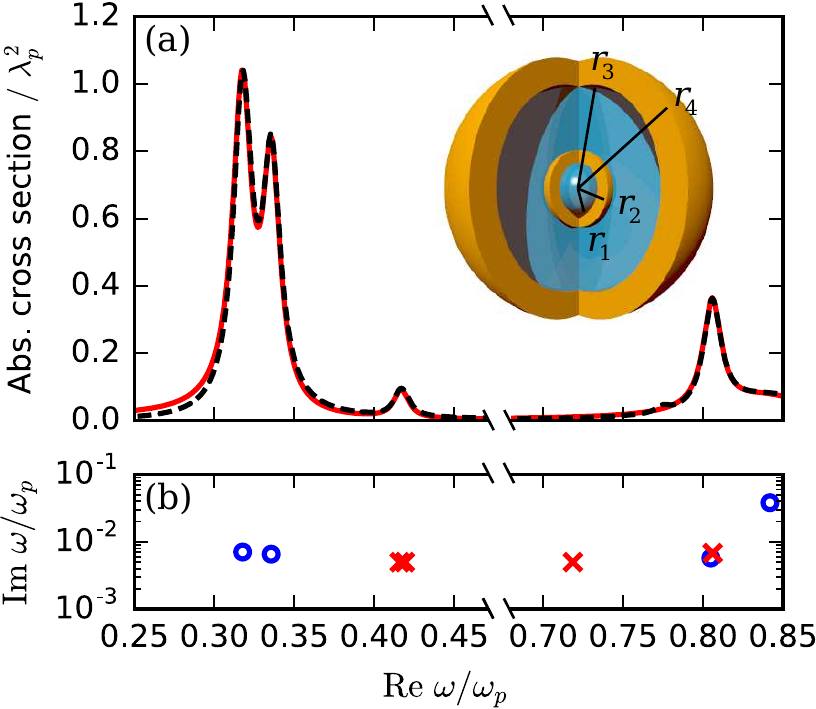}
\caption{(Color online) (a) Absorption cross section of a multilayered spherical nanoparticle constitued of a dielectric core and alternating layers of a Drude metal and a dielectric ($\eps = 2.1$), as shown in the inset. The red solid line is obtained from the quasinormal-mode expansion, whereas the dashed curve is the exact result from generalized Mie theory. The values of the radii of the different layers, starting with the inner one, are $r_1 = 0.012\lambda_p, r_2 = 0.0186\lambda_p, r_3 = 0.138\lambda_p, r_4 = 0.18 \lambda_p$, with all lengths being expressed in units of the plasma wavelength $\lambda_p = 2 \pi c / \w_p$. (b) Real and imaginary part of the quasinormal-mode eigenfrequencies included in the expansion. All modes are TM polarized. Circles and crosses refer to $l = 1$ and $l = 2$ modes, respectively, where $l$ is the azimuthal number.}\label{fig:particle} 
\end{figure}

In order to highlight the generality of the theory, we consider a very different example. As demonstrated in Refs.~\cite{soljacic2007,fan2012,soljacic2014}, coupled-mode theory can be used to model the scattering and absorption cross-sections of spatially confined scatterers, such as metallic nanoparticles. Although in these works only one or two quasinormal modes are included in the application of the theory, our formalism allows the extension of the number of modes and channels in a straightforward way. Moreover, we also use this example to illustrate the application of the theory to absorbing materials.

For three-dimensional scattering objects, the ports correspond to incoming and outgoing spherical waves of degree $l$, order $m$, and both transverse electric (TE) and transverse magnetic (TM) polarization \cite{bohren}. For simplicity, we consider a spherically symmetric system, where we can limit ourselves only to multipole terms with $m = 1$ and $l > 0$ \cite{bohren,soljacic2014}. The scattering and absorption cross section can be expressed as a function of the reflection coefficients, i.e., the diagonal terms of the scattering matrix, as follows \cite{soljacic_opt_expr}:
\begin{align}
\sigma_{\mathrm{sca}} &= \sum_{\sigma}\sum_{l=1}^{\infty}
\frac{\lambda^2}{8\pi} (2l + 1) \left| 1 - S_{l\sigma,l\sigma} \right|^2;\\
\sigma_{\mathrm{abs}} &= \sum_{\sigma}\sum_{l=1}^{\infty}
\frac{\lambda^2}{8\pi} (2l + 1) \left( 1 - \left|S_{l\sigma,l\sigma} \right|^2 \right)
\end{align}
(the index $\sigma$ indicates polarization: $\sigma =$ TE, TM). These expressions can be generalized to nonspherical scatterers by including the additional dependence on the order $m$ of the modes \cite{fan2012}.

For the sake of illustration, we consider a multilayered spherical particle with alternating layers of dielectric ($\varepsilon = 2.1$) and metallic materials, according to the structure sketched in the inset of Fig.~\ref{fig:particle}. Core--shell metallic nanoparticles are a viable and well established platform for obtaining a significant local field enhancement together with a broad frequency tunability in the spectral response \cite{halas}. Here, we are mainly interested in the  presence of multiple modes in each scattering channel, which underlines the advantages of our theoretical treatment in dealing with complex electromagnetic systems. Since we are considering a subwavelength particle sustaining plasmonic resonances, we limit ourselves to the lowest order TM-polarized modes ($l = 1$ and $l = 2$). We assume the metal dielectric function to follow Drude's model 
\begin{equation}
\varepsilon(\w) = 1 - \frac{\w_p^2}{\w(\w + i\kappa\nr)}
\end{equation}
with plasma frequency $\w_p$ and nonradiative damping rate $\kappa\nr$. In all calculations, we assume $\kappa\nr = 0.01 \w_p$. This value is in agreement with those obtained from the fitting of the dielectric function of noble metals (e.g., gold) at frequencies lower than the onset of interband transitions \cite{quasinormal_lalanne}. The complex eigenfrequencies of the modes have been extracted from the position of the poles of the exact reflection coefficient in the complex-frequency plane \cite{soljacic_opt_expr} and they are presented in Fig.~\ref{fig:particle}(b).

In the presence of absorbing materials, the theory requires the knowledge of the nonradiative decay rate of the modes, which is not directly available from our calculations, since the imaginary part of the complex eigenfrequency includes both the radiative and nonradiative components. In the case of low absorption, it is possible to distinguish the two contributions in an approximate way, by computing the complex eigenfrequencies twice, the second time upon setting Drude's damping rate to zero, and by taking the difference between the imaginary parts of the frequency in both calculations:
\begin{equation}
\gamma\nr[,j] = \Im\wt_j - \Im\wt_j^{(\kappa\nr = 0)}.
\end{equation}
The absorption cross section of the multilayered particles as calculated with our theory [Eqs.~\eqref{main3} and \eqref{Qnr}] is depicted in Fig.~\ref{fig:particle}(a) and it is compared with the exact result of generalized Mie theory \cite{bohren,soljacic_opt_expr}. The agreement of the curves is excellent, especially considering the additional level of approximation involved in estimating the nonradiative decay rates. Notable spectral features, such as the dip around $\omega = 0.33\omega_p$, which is due to the interference between partially overlapping $l = 1$ modes, or the significantly different oscillator strengths of $l = 1$ and $l = 2$ modes, are well reproduced by the scattering matrix expansion. These results demonstrate that the theory can be easily extended to non-unitary systems, when an estimate of the radiative efficiency of each quasinormal mode is available \cite{supplementary}.

\section{Discussion and conclusions}

In this work, we derived a general approach to expand the scattering matrix of optical systems on the basis of quasinormal modes and we validated it with illustrative examples. The theory is directly scalable to any number of modes and input--output channels. This particular feature allows us to treat all resonant modes on equal grounds, going beyond the traditional partition of a system in a small set of narrow modes and a frequency-dependent background fitted from simulation data. In this way, we achieve a more transparent picture of the modal structure of the system and, at the same time, we solve the ambiguity that could arise in defining the background channel in complex optical structures with a wide distribution of resonance widths. Eliminating the need for fitting a frequency-dependent background allows us to turn the qusinormal-mode expansion into a first-principle and self-consistent computational tool, which only requires the knowledge of the complex eigenfrequencies and the far-field behavior of the electromagnetic modes.

Creating artificial optical materials is an important goal in current nanophotonic research \cite{meta_review}. Such materials allow us to precisely control the intensity, phase, and polarization of scattered and transmitted light and to enhance light-matter interaction at the nanoscale. Spatial arrangements of optical resonators have been used, for instance, to realize high-contrast gratings \cite{krauss_hcg}, photonic metasurfaces \cite{capasso_review,brongersma,staude_aom}, and zero-refractive-index metamaterials \cite{zero-index}. When the constituting optical resonators are chiral, several intriguing effects can be observed, such as the asymmetric transmission of circularly and linearly polarized light \cite{fedotov2006,menzel2010}. Even for a single optical resonator, like a multilayered particle, the interference of different resonant states give rise to interesting phenomena, such as, for instance, the optical analog of electromagnetically induced transparency and superscattering \cite{fan2012_prl} and the formation of scattering dark states \cite{soljacic2014,alaee2015}. Multiple-resonance effects can also be exploited to tailor the scattering cross section of a scatterer, making it transparent to an outside observer \cite{alu}. Furthermore, hybrid photonic--plasmonic systems allow us to tailor the interaction with quantum emitters \cite{hybrid_chanda,hybrid_koenderink} and evidence polaritonic effects \cite{gomezrivas}. All these optical systems are typically characterized by a complex spectral structure, due to the presence of multiple electromagnetic modes coupled to the environment via various  incoming and outgoing channels.

Our theory establishes a direct connection between the electromagnetic modes and the spectral properties of photonic resonant systems. The expression for the quasinormal-mode expansion that we derive is reminescent of the Breit-Wigner formula \cite{nuclear,chaos}, albeit with the some notable distinctions. A crucial difference is that the coefficient of each resonant term in the expansion depends on the frequencies and the amplitudes of all the other modes via a specifically introduced coupling matrix $Q$. This additional dependence reflects the fact that, whereas the application of the Breit-Wigner formula is restricted to non-overlapping resonances, no such limitation applies to the present theory, which accounts in a natural way for the effective interaction among different states originating from the coupling to a common external environment.

Typically, as an alternative to modal expansion, the scattering matrix and the derived quantities (such as transmission or scattering intensities) can also be computed with a full-wave solver on a frequency-by-frequency basis. The expansion on quasinormal modes, however, offers several advantages over direct frequency-domain computations on several aspects. In the first place, modal methods allow for a significant reduction of computational times \cite{lalanne2,quasinormal_muljarov2}, especially when the presence of narrow resonances dictates a very fine frequency resolution. The most computationally demanding phase of the modal expansion is the calculation of the quasinormal modes. After that, the method allows us to arbitrarily reduce the frequency resolution at no further computational cost.

More importantly, the scattering matrix expansion provides a more complete amount of information and offers a deeper physical insight with respect to a frequency-by-frequency calculation. This aspect is especially helpful, for instance, in the process of designing and optimizing optical materials. Building upon the connection between quasinormal modes and scattering properties established by the theory, instead of looking for a specific spectral feature among a large number of simulated spectra with varying parameters, one could equivalently search for a quasinormal mode with specific attributes. This strategy is generally faster, more transparent, and more suggestive of the relation among the physical parameters. For all these reasons, the quasinormal-mode expansion of the scattering matrix is particularly suitable for investigating the physical mechanisms at the heart of highly structured spectra, such as those arising from the interference of several closely spaced modes. Indeed, as we noted above, this is the case for many photonic systems which are currently the subject of intense research efforts. At the same time, the theory also represents a powerful and predictive tool for the first-principle calculation of the scattering behavior of general physical systems.

\begin{acknowledgments}
This work is part of the research program of the Netherlands Organisation for Scientific Research (NWO). The authors acknowledge support from the European Research Council (ERC Advanced Grant 340438-CONSTANS) and from an industrial partnership program between Philips and NWO.
\end{acknowledgments}

\appendix

\section{Case of orthogonal modes}\label{nondiagonal}

If the scattering amplitudes of the quasinormal modes are orthogonal (i.e., $\vec{b}\dg_i\vec{b}_j = 0$ for $i \ne j$), or the spectral overlap between the modes can be neglected, the coupling matrix $Q$ of Eq.~\eqref{Q} becomes diagonal. The least-square solutions of Eq.~\eqref{main2} can, then, be written as $\lambda_j = - \vec{b}_j^T C\dg \vec{b}_j / (2\Im\,\tilde{\w}_j)$. As a result, the scattering-matrix expansion of Eq.~\eqref{main3} assumes the simpler expression:
\begin{equation}\label{single}
S = C + 2i\sum_{j=1}^{n} \frac{\Im\,\tilde{\w}_j}{\omega - \tilde{\omega_j}}
\frac{\vec{b}_j \vec{b}_j^T}{\vec{b}_j^T C\dg \vec{b}_j}.
\end{equation}
This equation can be understood as a modified version of the Breit-Wigner formula \cite{nuclear,chaos}, in which the interaction between overlapping modes is neglected, but where the relation between the phase of each resonant term and the direct-coupling matrix $C$ is retained. 

\section{Free choice of the normalization of the scattering amplitudes}\label{mult}

Here, we show that the result in Eq.~\eqref{main3} is independent of the normalization of the scattering amplitudes of the quasinormal modes. To this end, we consider a different set of amplitudes $\vec{b}'_j$, which differ from the original $\vec{b}_j$ by some complex multiplicative constants $\phi_j$ (which can be different for different modes):
\begin{equation}\label{sbar}
\vec{b}_j' = \phi_j \vec{b}_j.
\end{equation}
Introducing the diagonal matrix $\Phi = \mathrm{diag}(\phi_j)$ and the matrix $Q'$, defined by the expression in Eq.~\eqref{Q} with the modified eigenvectors, it is straightforward to verify that
$Q = \Phi^{*-1}\,Q'\,\Phi^{-1}$. In a similar fashion, we observe that Eq.~\eqref{main2} retains exactly the same form provided that $Q$, $B$, and $\Lambda$ are replaced by $Q'$, the column matrix of the new eigenvectors, $B'$, and $\Lambda' = \Phi \Lambda \Phi$, respectively. Then, substituting these replacements in Eq. \eqref{main}, we obtain that
\begin{equation}
S = C - i B \frac{1}{\omega\mathbb{I} -\tilde{\Omega}} \Lambda^{-1} B^T = C - i B' \frac{1}{\omega\mathbb{I} -\tilde{\Omega}} \Lambda'^{-1} B'^T,
\end{equation}
i.e., the expansion of the scattering maintains exactly the same formal expression independently of the choice of the eigenvector normalization constants.

\bibliography{bib}

%%%%%%%%% supplementary material %%%%%%%%%%%%%%%%%%%%
\onecolumngrid
\newpage
\pagestyle{empty}
\vspace*{\dimexpr-1in-\topmargin-\headsep-\headheight-\baselineskip}%
\hspace*{\dimexpr-1in-\evensidemargin-\parindent}%
{\includegraphics[page=1]{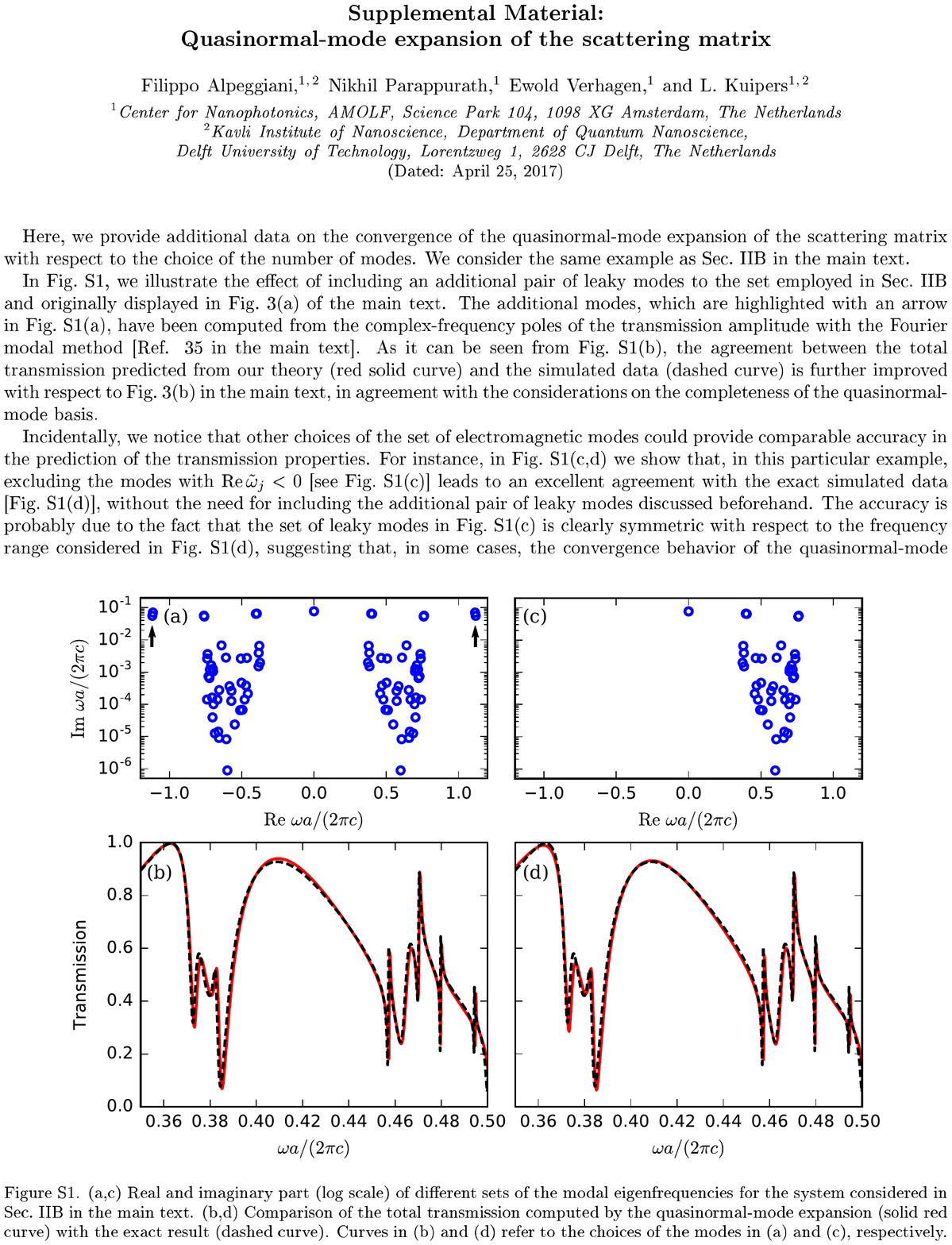}}

\newpage
\pagestyle{empty}
\vspace*{\dimexpr-1in-\topmargin-\headsep-\headheight-\baselineskip}%
\hspace*{\dimexpr-1in-\evensidemargin-\parindent}%
{\includegraphics[page=2]{supplementary}}

\newpage
\pagestyle{empty}
\vspace*{\dimexpr-1in-\topmargin-\headsep-\headheight-\baselineskip}%
\hspace*{\dimexpr-1in-\evensidemargin-\parindent}%
{\includegraphics[page=3]{supplementary}}

\end{document}